\documentclass[runningheads]{llncs}
\usepackage{amssymb}
\usepackage{amsmath}
\usepackage{amsfonts}
\usepackage{bm}
\usepackage{xspace}
\usepackage{graphicx}
\usepackage{array}
\usepackage{rotating}   
\usepackage{multirow}
\usepackage{makecell}
\usepackage{color}
\usepackage{colortbl}
\usepackage{enumerate}
\usepackage{booktabs}
\usepackage{subfig}
\usepackage{paralist}
\usepackage{pifont,tabularx,booktabs}
\usepackage{protocolj,arydshln}
\usepackage{array}
\usepackage{algorithm}
\usepackage{algpseudocode}
\usepackage{dashrule}

\newcolumntype{P}[1]{>{\centering\arraybackslash}p{#1}}

\newcommand{\Hzk}{\ensuremath{\mathsf{H}_{\mathsf{zk}}}}

\makeatletter
\g@addto@macro\normalsize{%
	\setlength\abovedisplayskip{2pt}
	\setlength\belowdisplayskip{2pt}
	\setlength\abovedisplayshortskip{2pt}
	\setlength\belowdisplayshortskip{2pt}
}
\makeatother

\usepackage[normalem]{ulem}

\usepackage{hyperref}
\usepackage[nameinlink,capitalize]{cleveref}

\renewcommand{\tablename}{\color{black}Table}
\renewcommand{\figurename}{\color{black}Figure}

\newcommand{\name}{SBvote\xspace}

\newcommand{\myparagraph}[1]{\vspace{0.1cm}\noindent{\it #1.}}

\begin{document}
\title{\name: Scalable Self-Tallying Blockchain-Based Voting}
%\thanks{}

\author{Ivana Stan\v{c}\'{i}kov\'{a} \and
Ivan Homoliak}
%Third Author\inst{3}\orcidID{2222--3333-4444-5555}}
%
\authorrunning{I. Stan\v{c}\'{i}kov\'{a} \and
	I. Homoliak}
\institute{Brno University of Technology, Czech Republic}
\maketitle              % typeset the header of the contribution
\begin{abstract}
Decentralized electronic voting solutions represent a promising advancement in electronic voting.
One of the e-voting paradigms, the self-tallying scheme, offers strong protection of the voters' privacy while making the whole voting process verifiable.
Decentralized smart contract platforms became interesting practical instantiation of the immutable bulletin board that this scheme requires to preserve its properties.
Existing smart contract-based approaches employing the self-tallying scheme (such as OVN or BBB-Voting) are only suitable for a boardroom voting scenario due to their scalability limitation.
The goal of our work is to build on existing solutions to achieve scalability without losing privacy guarantees and verifiability.
We present \name, a blockchain-based self-tallying voting protocol that is scalable in the number of voters, and therefore suitable for large-scale elections.
The evaluation of our proof-of-concept implementation shows that the protocol's scalability is limited only by the underlying blockchain platform.
We evaluated the scalability of \name on two public smart contract platforms -- Gnosis and Harmony.
Despite the limitations imposed by the throughput of blockchain platform, \name can accommodate elections with millions of voters.

\end{abstract}

\section{Introduction}
\label{sec:introduction}
Voting is an essential means of achieving a collective decision.
Traditionally, in large-scale voting such as national elections, the participants cast anonymous paper ballots that are later tallied by a trusted authority.
Following the advances in information technology, electronic voting systems have been introduced. 
Electronic voting takes place either in a controlled environment of a polling station via a dedicated machine or through the Internet via voters' own devices.

While several small-scale or boardroom e-voting protocols with decentralized architecture have been proposed~\cite{McCorrySH17,bbbvoting}, large-scale electronic voting systems mostly follow a centralized model~\cite{Adida08,cortier2019belenios}.
However, a centralized entity in control of the voting process represents a single point of failure as well as a possible element for misbehavior.
Therefore, we strive to leverage existing decentralized boardroom voting protocols and extend them to support large-scale voting.

The verifiability of many e-voting protocols depends on the assumed existence of a public bulletin board that should provide append-only modifications and immutability of the historical data~\cite{heather2008append}.
All participants can read the data on the public bulletin board and use them to verify the correctness of the voting process.
Voting systems such as Helios~\cite{Adida08} implement the public bulletin board as a single web server.
This, however, introduces the possibility of several issues, including unavailability of the server (e.g., due to a denial-of-service attack) or censorship by a malicious authority controlling the server.

Other systems~\cite{McCorrySH17,bbbvoting,yang2021priscore} instantiate the public bulletin board on a blockchain with a smart contract platform.
On top of the immutability and append-only features, such blockchains also provide correct execution of a code that enables decentralized e-voting schemes to utilize public verifiability of the data submitted to the bulletin board (e.g., votes and a tally). 
Protocols such as Open Vote Network (OVN)~\cite{McCorrySH17} and BBB-Voting~\cite{bbbvoting} use smart contracts to orchestrate the procedures of the boardroom voting protocol.
The distributed nature of these protocols also eliminates the need to rely on the authority to tally the votes. 
In these approaches, referred to as self-tallying voting, any participant can perform and verify the tally computation.

Since BBB-Voting, in contrast to OVN, provides more than two voting choices and enables the recovery of faulty participants, we base our work on BBB-Voting.
In detail, our goal is to build a voting protocol that resolves the scalability limitation of BBB-Voting while maintaining maximum voter privacy.
We introduce \name, a decentralized blockchain-based voting protocol that provides scalability in the number of participants by grouping them into voting booths instantiated as dedicated smart contracts that are controlled and verified by the aggregation smart contract.
As a result, our approach is suitable for privacy-preserving self-tallying large-scale e-voting.

\myparagraph{\textbf{Contributions}}
We make the following contributions:

\begin{compactenum}[i)]
	 \item 
	 We introduce SBvote as an extension of the BBB-Voting protocol, which enables scalability in the number of voters.
	 SBvote introduces multiple voting smart contracts booths that are managed and aggregated by a main smart contract. 
	 
	 \item Our extended solution maintains all properties of decentralized e-voting, including public verifiability, perfect ballot secrecy, and fault tolerance. Moreover, it improves the privacy of voters within booths.

     \item We made a proof-of-concept implementation and evaluated it on two smart contract platforms, Harmony and Gnosis (formerly known as xDai).
	 We achieved the best scalability results using the Harmony blockchain, allowing us to run elections with 1.5M voters and two candidates within a two-days interval.

\end{compactenum}

\myparagraph{\textbf{Organization}}
The rest of the paper is organized as follows.
Essential preliminaries are presented in \autoref{sec:preliminaries}.
We introduce \name in \autoref{sec:protocol} and evaluate its scalability in \autoref{sec:evaluation}.
We provide the security analysis of \name and discuss its properties in \autoref{sec:discussion}.
We review the related work in \autoref{sec:relatedwork} and finally conclude our paper in \autoref{sec:conclusion}.

\section{Preliminaries}
\label{sec:preliminaries}
This section briefly reviews the properties of voting protocols and provides a short description of blockchains and smart contracts.
Finally, we describe the BBB-Voting protocol on which we base our work.

\subsection{Voting}
\label{ssec:evoting}
We provide a list of several desired properties of voting protocols in the following.

%\begin{compactitem}
		\noindent\textbf{Privacy:} 	
		The votes remain anonymous.
		Only the voter herself knows which candidate she voted for.
		Privacy protection is a crucial attribute of voting systems used in practice and is very important in publicly verifiable voting schemes.
		
		\noindent\textbf{Perfect Ballot Secrecy:} 
		was defined by Kiayias and Yung~\cite{Kiayias2002}, and it extends the privacy property. In a scheme with perfect ballot secrecy, a partial result can only be revealed if all remaining voters collude to uncover it.
		
		\noindent\textbf{Self-Tallying:}
		ensures that any interested party can compute the tally without unblinding the individual votes.
		
		\noindent\textbf{Fault Tolerance:}
		The protocol allows for excluding faulty participants in a publicly verifiable manner without restarting the whole voting protocol. 
		
		\noindent\textbf{Verifiability:}
		ensures that the correctness of the voting process can be checked.
		Verifiability includes individual verifiability (allowing the voter to verify her vote has been counted) and universal verifiability (allowing any interested party to verify all cast votes have been correctly tallied).
		Furthermore, the verifiability of a voting system can be described as follows~\cite{benaloh2015end}: 
		
		\begin{compactitem}
			\item \textbf{cast-as-intended}: a voter can verify the encrypted vote contains her choice of candidate,
			\item \textbf{recorded-as-cast}: a voter can verify the system recorded her vote correctly,
			\item \textbf{tallied-as-recorded}: any interested party is able to verify whether the final tally corresponds to the recorded votes.
		\end{compactitem}
	
		A voting system satisfying all these properties is end-to-end verifiable.
		
		\noindent\textbf{Dispute-Freeness:}
		The protocol's design prevents disputes among involved parties by allowing anyone to verify whether a participant followed the protocol.

		\noindent\textbf{Completeness:}
		All valid votes are included in the final tally.

\subsection{Blockchains and Smart Contracts}
\label{ssec:blockchains}
Blockchain is a continuously growing distributed database consisting of blocks maintained by a network of consensus nodes (i.e., that run a consensus protocol).
Once the consensus nodes agreement on a new block, it is added to the blockchain.
The blocks are cryptographically linked to ensure the immutability of the entire ledger, and they contain records of cryptocurrency transfers executed within the network.

Orders to execute transfers are communicated to the network in messages called \emph{transactions}.
The block may also contain application code written in a supported language on blockchain platforms that support smart contracts.
This code is invoked by a transaction containing execution orders (i.e., function calls of a smart contract).
The blockchain network then acts as a decentralized computation platform -- the blockchain nodes execute the smart contract code.

On many smart contract platforms (such as Ethereum~\cite{wood2014}), the execution complexity of smart contract invocations is measured in units of gas.
The sender of a transaction containing a smart contract function call has to pay for the consumed gas to cover the expenditure made by the nodes for carrying out the computation.
The gas price is volatile and based on the demand on the network.

\subsection{BBB-Voting}\label{sec:prelim-bbb}
BBB-Voting~\cite{bbbvoting} is a system for boardroom voting supporting $k \geq 2$ choices. 
The basic protocol used in BBB-Voting consists of five phases (i.e., registration, setup, pre-voting, voting, tally) and an optional fault-recovery phase.
In the first phase, a voting authority registers eligible voters and their wallet addresses. 
In the second phase, common generator $g\in \mathbb{F}_p^*$ is chosen, where $p=2\cdot q +1$ and $q$ is a prime. 
The number of participants is denoted as $n$ and $n < p - 1$.
To represent $k$ candidates, independent generators $\{f_1,...,f_k\}$ in $\mathbb{F}_p^*$ are selected.
Each voter $P_i$ then creates her ephemeral private key as $x_i \in_{R} \mathbb{F}^{\ast}_p$ and ephemeral public key as $g^{x_i}$.
All ephemeral public keys are submitted to a smart contract.
The multi-party computation (MPC) key for each participant is computed by the smart contract in the third phase as follows:
	\begin{equation}
	h=g^{y_i}=\prod\limits_{j=1}^{i-1} g^{x_j}/\prod\limits_{j=i+1}^{n} g^{x_j},
	\end{equation}
where $y_i=\sum_{j<i}x_j- \sum_{j>i}x_j$ and $\sum_{i}x_i y_i=0$.
In the fourth phase, voter $P_i$ computes her blinding key $g^{x_i y_i}$ and then submits her blinded vote $B_i = g^{x_i y_i} f_j$ to the smart contract, where $f_j \in {f_1, ..., f_k}$ denotes the selected candidate.
A 1-out-of-$k$ non-interactive zero-knowledge (NIZK) proof of set membership is submitted to the smart contract along with the blinded vote.
Next, the smart contract verifies the correctness of such a vote.
During the fifth phase of the protocol, the tally of votes is computed off-chain by an arbitrary party and submitted to the smart contract.
The smart contract verifies that the tally satisfies the following equation:
	\begin{equation}
	\prod\limits_{i=1}^{n}B_i=
	\prod\limits_{i=1}^{n}g^{x_iy_i}f=
	g^{\sum_{i}x_i y_i}f=
	{f_1}^{ct_1}{f_2}^{ct_2}...{f_k}^{ct_k}.
	\end{equation}
The protocol also includes an optional fault-recovery extension that can be placed after the voting phase. 
This phase is useful if one or more participants have stalled and have not cast their blinded votes.

The BBB-Voting scheme provides perfect ballot secrecy, fairness, public verifiability, self-tallying feature, dispute-freeness, resistance to serious failures, and maximizes the voters’ privacy (see \autoref{ssec:evoting}).
Also, it introduces several optimizations of the implementation to decrease the costs of the protocol and accommodate a larger number of participants than the previous approaches  (i.e., OVN~\cite{McCorrySH17}).

BBB-Voting is designed as a single smart contract on the Ethereum block\-chain.
Nevertheless, BBB-Voting is intended only for boardroom voting with a low number of involved participants.
Hence it does not provide scalability as might be required in national elections.
Another limitation of BBB-Voting is the low number of stalling participants the system can recover from in a single fault recovery round due to the block gas limit.

\section{Scalable Voting Protocol}
\label{sec:protocol}
In this section, we propose \name, a scalable e-voting protocol that is based on BBB-Voting (see~\autoref{sec:prelim-bbb}). 

\subsection{System Model}
\label{ssec:sys-model}
We focus on a decentralized e-voting that provides all desired properties of e-voting schemes mentioned in \autoref{ssec:evoting} and provides scalability in the number of the participants.
We also assume a centralized authority that is responsible for the enrollment of the participants and shifting the stages of the protocol. 
However, the authority can neither change nor censor the votes of the participants, and it cannot compromise the privacy of the votes.

We assume that a public bulleting board required for e-voting is instantiated by a blockchain platform that moreover supports the execution of smart contracts. 
We assume that all participants of voting have their thin clients that can verify the inclusion of their transactions in the blockchain as well as the correct execution of the smart contract code.

\myparagraph{\textbf{Adversary Model}}
We consider an adversary that passively listens to communication on the blockchain network.
She cannot block, modify, or replace any transaction on the network.
The adversary can link a voter's IP address to her blockchain address.
However, she does not possess the computational resources to break the cryptographic primitives used in the blockchain platform and the voting protocol.
The adversary cannot access or compromise the voter's device or the user interface of the voting application.
We assume that in each voting group of $n$ participants, at most $t$ of them can be controlled by the adversary and disobey the voting protocol, where $t \leq n - 2$ and $n \geq 3$.

\subsection{Proposed Approach}
\myparagraph{\textbf{Involved Parties}}
Our proposed approach has the following actors and components: (1) \textit{a participant $\mathbb{P}$} (also referred to as \textit{a voter}) who chooses a candidate and casts a vote,
(2) \textit{a voting authority $\mathbb{VA}$} who is responsible for the registration of participants and initiating actions performed by smart contracts,
(3) \textit{the main contract $\mathbb{MC}$}, which assigns participants to voting booths, deploys booth contracts, and aggregates the final tally from booth contracts,
(4) \textit{a booth contract $\mathbb{BC}$}, which is replicated into multiple instances while each instance serves a limited number of participants. 
New instances might be added on-demand to provide scalability.

\begin{figure}[p]
	\centering
	\includegraphics[width=0.99\linewidth,height=0.99\textheight,keepaspectratio]{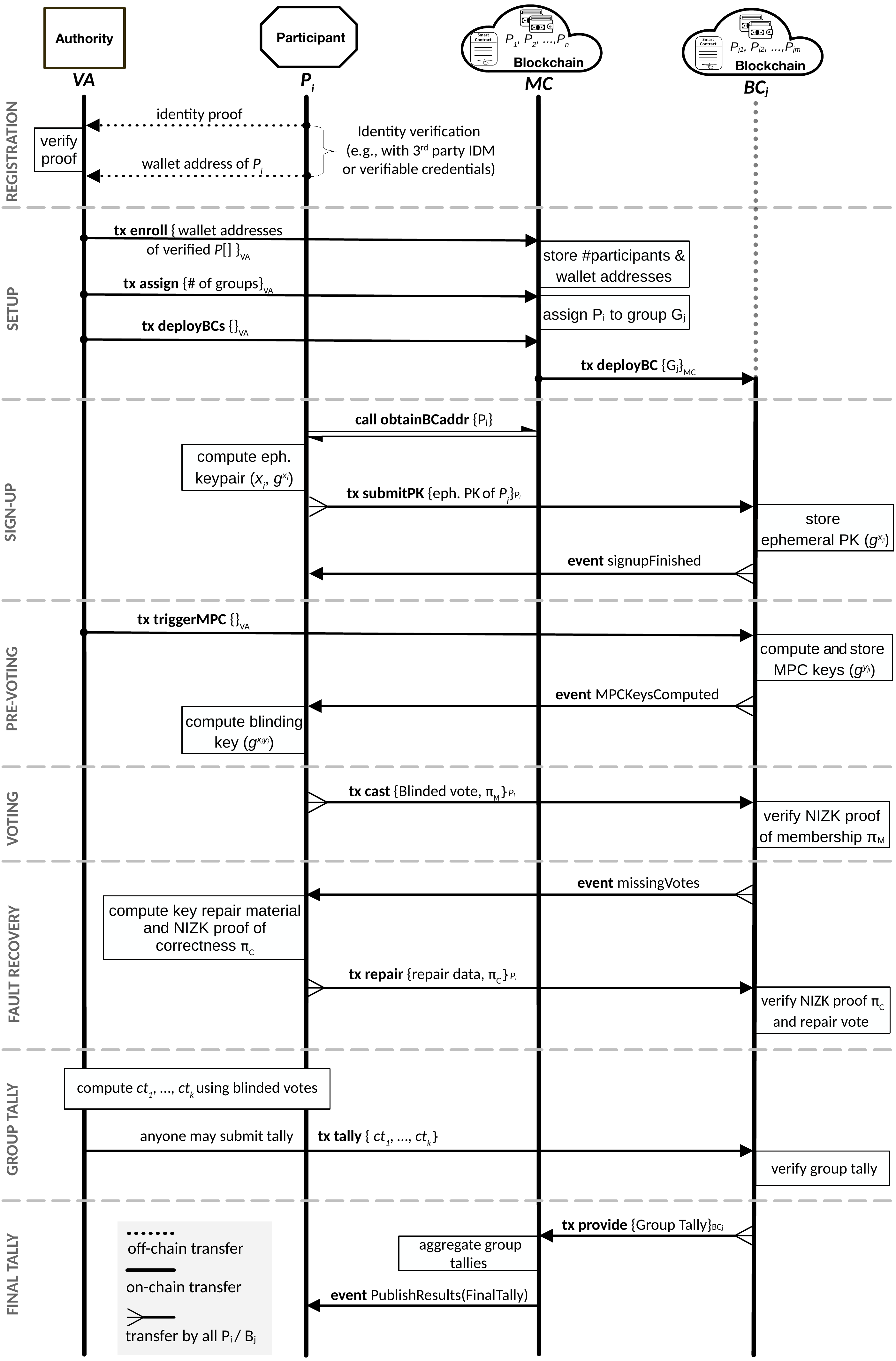}
	\caption{Overview of \name protocol.}
	\label{fig:operation-scheme-bw}
\end{figure}

\myparagraph{\textbf{Protocol}}
We depict our protocol in \autoref{fig:operation-scheme-bw}. \name follows similar phases as BBB-Voting but with several alterations that enable better scalability.
Just like in BBB-Voting, the registration phase assumes the existence of an Identity Management (IDM) system that allows $\mathbb{VA}$ to authenticate users and generate a list of eligible voters. 
In BBB-Voting, the setup phase of the protocol allows users to submit their ephemeral public keys. 
However, in contrast to BBB-Voting, \name requires additional steps to set up the booth contracts. 
First, eligible voters are assigned to voting groups, and then $\mathbb{BC}$ is deployed for each voting group.
Once the setup is finished, voters proceed to submit their ephemeral public keys during a sign-up phase.
These keys are further used to compute multi-party computation (MPC) keys within each voting group during a pre-voting phase.
In the voting phase, voters cast their blinded votes along with corresponding NIZK proofs.
The NIZK proof allows the smart contract to verify that the blinded vote correctly encrypts one of the valid candidates.
If some of the voters who submitted their ephemeral public keys have failed to cast their vote, the remaining active voters repair their votes in the subsequent fault recovery phase.
This is achieved by removing the stalling voters' key material from the encryption of the correctly cast votes.
The key material has to be provided by each active voter along with NIZK proof of correctness.
After the votes are repaired, the tallies for individual voting groups are computed during a booth tally phase.
Then, partial results are aggregated to obtain the final tally.

In the following, we describe the phases of our protocol in more detail.
Phases 2--6 are executed in parallel in each of the voting groups/booth contracts. \\

\myparagraph{\textbf{Registration}}
\label{ssec:registration}
In this phase, the participants interact with $\mathbb{VA}$ to register as eligible voters for the voting. 
A suitable IDM system is required, allowing the $\mathbb{VA}$ to verify participants' identities and eligibility to participate in the voting.
Each participant creates her blockchain wallet address and registers it with the $\mathbb{VA}$. 
The $\mathbb{VA}$ stores a mapping between a participant's identity and her wallet address. 

\myparagraph{\textbf{Phase 1 (Setup)}}
\label{ssec:setup}
First, the voting authority deploys the main contract to the blockchain.
Then, $\mathbb{VA}$ enrolls the wallet addresses of all registered participants to $\mathbb{MC}$ within a transaction.\footnote{Note that in practice this step utilizes transaction batching to cope with the limits of the blockchain platform (see \autoref{ssec:optim}).}
Once all the registered participants have been enrolled, $\mathbb{VA}$ triggers $\mathbb{MC}$ to pseudo-randomly\footnote{Note that distributed randomness protocols such as RoundHound~\cite{syta2017scalable} might be used for this purpose, however, in this work we assume a trusted randomness source that is agreed upon by all voters (e.g., a hash of some Bitcoin block). } distribute enrolled participants into groups whose size is pre-determined and ensures a certain degree of privacy. 

In every group, the participants agree on the parameters of the voting.
Let $n$ be the number of participants in the group and $k$ the number of candidates.
We specify the parameters of voting as follows:

\setlength{\parskip}{5pt} \setlength{\itemsep}{5pt}
\begin{compactenum}[$\;$]
	\item[1)] a common generator $g\in \mathbb{F}_p^*$, where $p=2\cdot  q +1$, $q$ is a prime and $n < p - 1$.
	
	\item[2)] $k$ independent generators $\{f_1,...,f_k\}$  in $\mathbb{F}_p^*$ such that $f_i = g^{2^{(i - 1)m}}$, where $m$ is the smallest integer such that $2^m > n$.
\end{compactenum}

Then, $\mathbb{VA}$ deploys a booth contract $\mathbb{BC}$ for each group of participants with these previously agreed upon voting parameters.
$\mathbb{MC}$ stores a mapping between a participant's wallet address and the group she was assigned to.

\myparagraph{\textbf{Phase 2 (Sign-Up)}}
\label{ssec:signup}
Eligible voters enrolled in the setup phase review the candidates and the voting parameters.
Each voter who intends to participate obtains the address of the $\mathbb{BC}$ she was assigned to by $\mathbb{MC}$. 
From this point onward, each participant interacts only with her $\mathbb{BC}$ representing the group she is part of. 
Every participant $P_i$ creates her ephemeral key pair consisting of a private key $x_i \in_{R} \mathbb{F}^{\ast}_p$ and public key $g^{x_i}$.
The $P_i$ then sends her public key to $\mathbb{BC}$. 
By submitting an ephemeral public key, the participant commits to cast a vote later.  
Furthermore, participants are required to send a deposit within this transaction.
If the voter does not cast her vote or later does not participate in the potential fault recovery phase, she will be penalized by losing the deposit.
Voters who participate correctly retrieve their deposit at the end of the voting.

\myparagraph{\textbf{Phase 3 (Pre-Voting)}}
\label{ssec:prevoting}
In this step, each $\mathbb{BC}$ computes synchronized multi-party computation (MPC) keys from the participants' ephemeral public keys submitted in the previous step. 
To achieve scalability, the MPC keys are computed independently in each $\mathbb{BC}$ over the set of ephemeral public keys within the group.
The MPC key for participant $P_i$ is computed as follows:
	\begin{equation}
	\label{eqn:eqn1}
	h=g^{y_i}=\prod\limits_{j=1}^{i-1} g^{x_j}/\prod\limits_{j=i+1}^{n} g^{x_j},
	\end{equation}
	where $y_i=\sum_{j<i}x_j- \sum_{j>i}x_j$
	and
	$\sum_{i}x_i y_i=0$ (see Hao et al.~\cite{HaoRZ10} for the proof).
	
The computation of MPC keys is triggered by $\mathbb{VA}$ in each $\mathbb{BC}$.
After the computation, each participant obtains her MPC key from $\mathbb{BC}$ and proceeds to compute her ephemeral blinding key as $g^{x_i y_i}$ using her private key $x_i$.

\myparagraph{\textbf{Phase 4 (Voting)}}
\label{ssec:voting}
Before participating in this phase of the protocol, each voter must create her blinded vote and a NIZK proof of its correctness.
The blinded vote of the participant $P_i$ is $B_i = g^{x_i y_i} f_j$, where $f_j \in {f_1, ..., f_k}$ represents her choice of a candidate.
The participant casts the blinded vote by sending it to $\mathbb{BC}$ in a transaction $cast(B_{i}, \pi_M)$, where $\pi_M$ is a 1-out-of-$k$ NIZK proof of set membership. 
This proof allows the $\mathbb{BC}$ to verify that the vote contains one of the candidate generators from ${f_1, ..., f_k}$ without revealing the voter's choice.
$\mathbb{BC}$ performs a check of the proof's correctness and accepts well-formed votes.
Construction and verification of the NIZK proof are depicted in \autoref{fig:ZKPMultiCandidate}.

\begin{figure}[t]
	\begin{center}
		\tiny
		\fbox{
			\begin{protocolm}{2}						
				\participants{\underline{Participant $P_i$}}{\underline{Smart Contract}}
				\participants{($~h\leftarrow g^{y_i},~ v_i$)}{($~h\leftarrow g^{y_i}$)}
				\hline	
				& &\\		
				%				\text{Generates~vote~choice~of} \\
				Select~v_i\in\{1,...,k\}, & & \\	 				
				Use~choice~generators~\\
				f_l\in\{f_1,...,f_k\}\subseteq\mathbb{F}_p^*,\\			
				%Publish:~B_i\leftarrow h^{x_i}{f_l}\\
				Publish~ x\leftarrow g^{x_i} && 	\\
				Publish~ B_i\leftarrow h^{x_i}{f_l} & & \\
				
				%h\leftarrow g^{y_i} && h\leftarrow g^{y_i}(Computed~in~Equation~ \ref{eqn:eq1})	\\
				
				w\in_{R} \mathbb{F}_p^*\\
				%a_{v_j}\leftarrow g^{w} \\
				%b_{v_j}\leftarrow h^{w} \\

				\forall l\in \{1,..,k\}\setminus {v_i}:\\
				
				\quad 1.~r_l,d_l\in_{R} \mathbb{F}_p^* \\
				\quad 2.~a_l\leftarrow x^{-d_l}g^{r_l}\\ 				
				\quad 3.~b_l\leftarrow h^{r_l}(\frac{B_i}{f_l})^{-d_l} \\

				for\enspace {v_i}:\\
				
				\quad 1.~a_{v_i}\leftarrow g^{w}\\
				\quad 2.~b_{v_i}\leftarrow h^{w} \\
				%&\stackrel{\forall l: \{a_l\},\{b_l\}}{\xrightarrow{\hspace*{4.5cm}}}&\\
				\\%\hdashline
				c\gets\Hzk(\{ \{a_l\},\{b_l\} \}_{l})\\
				\\%\hdashline
				%&\sends{\forall l: \{a_l\},\{b_l\}}&\\
				%&\receives{c}&c\leftarrow \mathbb{Z}_q  \\
				
				for\enspace {v_i}:\\
				
				%\quad 1.~d_{v_i}\leftarrow c \bigoplus_{l\neq{v_i}}d_l\\
				\quad 1.~d_{v_i}\leftarrow \sum_{l\neq{v_i}}d_l\\
				\quad 2.~d_{v_i}\leftarrow c - d_{v_i}\\
				
				\quad 3.~r_{v_i}\leftarrow w+x_id_{v_i} \\			
				\quad 4.~q\leftarrow p-1\\
				\quad 5.~{r_{v_i}\leftarrow r_{v_i}\mod{q}}\\

				%&\stackrel{\forall l:\{r_l\},\{d_l\}}{\xrightarrow{\hspace*{4.5cm}}}
				&\forall l: \{a_l\}, \{b_l\}, &\\
				&\sends{\{r_l\},\{d_l\}}
				&\\
				
				& &\Psi \gets \{\forall l: \{a_l\},\{b_l\}\}  \\
				&& c \gets\Hzk(\Psi)\\
				
				%& &d_{sum}\leftarrow\sum_{l}d_l~mod~p-1\\
				& &\sum_{l} d_{l} \stackrel{?}{=} c\\
				
				&& \forall l\in \{1,..,k\}\\
				%&& && x^{{r_{v_j}}}\stackrel{?}{=}{a_{v_j}}x^{d_{v_j}}\\

				&&\quad 1.~g^{r_l}\stackrel{?}{=}{a_l}x^{d_l}\\
				&&\quad 2.~h^{r_l}\stackrel{?}{=}{b_l}(\frac{B_i}{{f_l}})^{d_l}\\
				%h^{r_{v_j}}\stackrel{?}{=}{b_{v_j}}(B_i/{f_l})^{d_{v_j}}\\			
			\end{protocolm}
		}
	\end{center}
	\vspace{-0.35cm}
	\caption{Non-interactive zero-knowledge proof of 1-out-of-$k$ set membership.}
	\label{fig:ZKPMultiCandidate}
\end{figure}	

\begin{figure}[t]
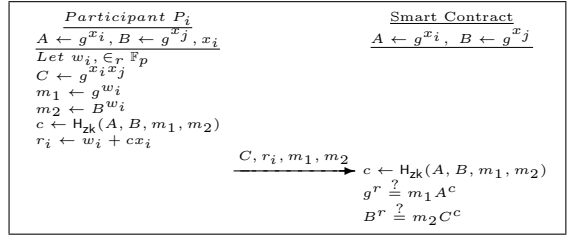

	\begin{center}
		\fbox{
			\tiny
			\begin{protocolm}{2}	
				
				\participants{\underline{$Participant~P_i$}}{\underline{Smart Contract}}%\\ 						
				\participants{\underline{$A\leftarrow g^{x_i},B
						\leftarrow g^{x_j},x_i$}}{\underline{$A\leftarrow g^{x_i},~B\leftarrow g^{x_j}$}}%\\				
				Let~w_i,\in_{r}\mathbb{F}_p\\
				C\leftarrow g^{x_ix_j}\\
				m_1\leftarrow {g}^{w_i}\\
				m_2\leftarrow {B}^{w_i}  \\ %&&\stackrel{m_1,m_2}{\xrightarrow{\hspace*{0.75cm}}}\\

				c\leftarrow  \Hzk(A,B,m_1,m_2)\\ 	
				r_i\leftarrow w_i+ cx_i &&\\
				&\sends{C,r_i,m_1,m_2} & c\leftarrow  \Hzk(A,B,m_1,m_2)\\			
				%&& &&(From~reconstructed~s_{ij})\\

				&  & g^r\stackrel{?}{=}{m_1}{A^c}\\
				&  & B^r\stackrel{?}{=}{m_2}{C^c}\\

			\end{protocolm}
		}
	\end{center}
	\vspace{-0.35cm}
	\caption{NIZK proof verifying correspondence of $g^{x_ix_j}$ to public keys $A=g^{x_i}, B=g^{x_j}$.}
	\label{fig:ZKPdiffie}
\end{figure}

\myparagraph{\textbf{Phase~5~(Fault-Recovery)}}
\label{ssec:faultrec}
The use of synchronized MPC keys ensures a vote cast by each voter contains a key material shared with every other voter within the group.
If some of the voters within a group stall during the voting phase, the tally cannot be computed from the remaining data.
Therefore, we include a fault-recovery phase, where remaining voters provide $\mathbb{BC}$ with the key material they share with each stalling voter, enabling $\mathbb{BC}$ to repair their votes.
In detail, for a stalling voter $P_j$ and an active voter $P_i$ ($i \neq j$), the shared key material $g^{x_i x_j}$ consists of the stalling voter's ephemeral public key $g^{x_j}$ (previously published in $\mathbb{BC}$) and the active voter's ephemeral private key $x_i$. 
The active voters send the shared key material to $\mathbb{BC}$ along with a NIZK proof depicted in \autoref{fig:ZKPdiffie}. The NIZK proof allows the $\mathbb{BC}$ to verify that the shared key material provided by the voter corresponds to the ephemeral public keys $g^{x_i}$ and $g^{x_j}$.

Suppose some of the previously active voters become inactive during the fault-recovery phase (i.e., do not provide the shared key material needed to repair their votes).
In that case, the fault-recovery phase can be repeated to exclude these voters.
Note that this phase takes place in groups where all the voters who committed to vote during the sign-up phase have cast their votes.

\myparagraph{\textbf{Phase~6~(Booth Tallies)}}
\label{ssec:boothtally}
At first, the tally has to be computed for each group separately.
Computation of the result is not performed by the $\mathbb{BC}$ itself.
Instead, $\mathbb{VA}$ (or any participant) obtains the blinded votes from the $\mathbb{BC}$, computes the tally, and then sends the result back to $\mathbb{BC}$.
$\mathbb{BC}$ verifies whether a provided tally~fits \autoref{eqn:eq3}.
\begin{equation}
	\label{eqn:eq3}
	\prod\limits_{i=1}^{n}B_i=
	\prod\limits_{i=1}^{n}g^{x_iy_i}f=
	g^{\sum_{i}x_i y_i}f=
	{f_1}^{ct_1}{f_2}^{ct_2}...{f_k}^{ct_k}.
\end{equation}
\myparagraph{\textbf{Phase~7~(Final Tally)}}
\label{ssec:finaltally}
Once $\mathbb{BC}$ obtains a correctly computed tally, it sends it to $\mathbb{MC}$. 
$\mathbb{MC}$ collects and summarizes the partial tallies from individual booths and announces the final tally once all booths have provided their results. 
The participants can also review the partial results from already processed booths without waiting for the final tally since the booths tallies are processed independently.

\subsection{Design Choices and Optimizations}
\label{ssec:optim}
We introduce several specific features of \name, which allow us to achieve the scalability and privacy properties.

\myparagraph{\textbf{Storage of Voters' Addresses}}
If we were to store the voters' wallet addresses in the booth contracts, it would cause high storage overhead and thus high costs.
However, we proposed to store these addresses only in $\mathbb{MC}$, while booth contracts can only query $\mathbb{MC}$ whenever they require these addresses (i.e., when they verify whether a voter belongs to the booth's group).
As a result, this eliminates the costs of transactions when deploying booth contracts, saving the blockchain storage space.

\myparagraph{\textbf{Elimination of Bottlenecks}}
The main focus of our proposed approach is the elimination of the bottlenecks that limit the total number of voters and the size of the voting groups. 
In particular, passing the data within a single transaction could potentially exceed the block gas limit.

The scalability of the Setup phase of \name is straightforward to resolve since it does not involve any transient integrity violation checks (excluding checking for duplicates).
In all these cases, $\mathbb{VA}$ splits the data into multiple independent transactions.
Similarly, each active voter can send the key material required to repair her vote in several batches in the Fault-Recovery phase, allowing the system to recover from an arbitrary number of stalling participants.

In contrast to the Setup and Fault-Recovery phases, batching in the Pre-Voting phase is not trivial since it requires transient preservation of integrity between consecutive batches of the particular voting group.
Therefore, we used a custom batching mechanism, which eliminates this bottleneck while also optimizing the cost of the MPC computation.

\myparagraph{\textbf{MPC Batching and Optimization}}
If computed independently for each participant, the computation of MPC keys leads to a high number of overlapping multiplications.
Therefore, we optimize this step by dividing the computation into two parts, respecting both sides of the expression in \autoref{eqn:eqn1} and reusing accumulated values for each side.

First, we pre-compute the right part (i.e., divisor) of \autoref{eqn:eqn1}, which consists of a product of ephemeral public keys of voters with a higher index than the current voter's one (i.e., $i$ in \autoref{eqn:eqn1}).
The product is accumulated and saved in the contract's storage at regular intervals during a single iteration over all ephemeral public keys.
The size of these intervals corresponds to the batch size chosen for the computation of the remaining (left side) of the equation.
We refer to these saved values as \emph{right markers} (see \autoref{alg:precomp}).
We only choose to save the right markers instead of saving all accumulated values due to the high cost of storing data in the contract's persistent storage.
Though the intermediate values between right markers have to be computed again later, they are only kept in memory (not persistent between consecutive function calls).
Therefore, they do not significantly impact the cost of the computation.

\begin{algorithm}[t]
	\scriptsize
	\caption{Pre-computation of right side values of \autoref{eqn:eqn1}}
	\label{alg:precomp}
	\textbf{Inputs:}
	\vspace{-8pt}
	\begin{itemize}
		\setlength\itemsep{0.1em}
		\item the number of voters $n$
		\item batch size for MPC computation $mpc\_batch$
		\item array of voters' ephemeral public keys $voterPKs$
	\end{itemize}
	\vspace{-8pt}
	\textbf{Outputs:}
	\vspace{-8pt}
	\begin{itemize}
		\setlength\itemsep{0.1em}
		\item pre-computed right side values $right\_markers$
	\end{itemize}
	\vspace{-13pt}
	\hrulefill
	\begin{algorithmic}[1]
		\State $right\_tmp \gets 0$
		\If{$n \;\mathrm{mod}\; mpc\_batch \neq 0$}
		\State $right\_markers.$push($right\_tmp$)
		\EndIf
		\For{$i \gets 0$ to $n$}
		\If{$n \;\mathrm{mod}\; mpc\_batch$ = $(i-1) \;\mathrm{mod}\; mpc\_batch$}
		\State $right\_markers.$push($right\_tmp$)
		\EndIf
		\State $right\_tmp \gets right\_tmp \times voterPKs[n - i]$
		\EndFor
	\end{algorithmic}
\end{algorithm}

The second part of the computation is processed in batches.
First, the right-side values for all voters within the current batch are obtained using the pre-computed right marker corresponding to this batch (see lines 1--5 of \autoref{alg:mpc}).
Then,  the left part of \autoref{eqn:eqn1} is computed for each voter within the batch, followed by evaluating the entire equation to obtain the MPC key (lines 6--9 of \autoref{alg:mpc}).
This left-side value is not discarded; therefore, computing the left side for the next voter's MPC key only requires single multiplication.
The last dividend value in the current batch is saved in the contract's storage to allow its reuse for the next batch.

\begin{algorithm}[t]
	\scriptsize
	\caption{Computation of a batch of MPC keys}
	\label{alg:mpc}
	\textbf{Inputs:}
	\vspace{-8pt}
	\begin{itemize}
		\setlength\itemsep{0.1em}
		\item array of voters' ephemeral public keys $voterPKs$
		\item batch size for MPC computation $mpc\_batch$
		\item start index $start$ and the end index $end$ of the current batch
		\item pre-computed right side value for the first index in batch $right\_marker$
		\item left side value from the previous batch $act\_left$
	\end{itemize}
	\vspace{-8pt}
	\textbf{Outputs:}
	\vspace{-8pt}
	\begin{itemize}
		\setlength\itemsep{0.1em}
		\item left side value at the last index of the current batch $act\_left$
		\item array of MPC keys for the current batch $mpc\_keys$
	\end{itemize}
	\vspace{-13pt}
	\hrulefill \\
	Compute right side values for the batch:
	\begin{algorithmic}[1]
		\State $right\_tab[mpc\_batch - 1] \gets right\_marker$
		\For{$i \gets 0$ to $mpc\_batch$}
		\State $j \gets mpc\_batch - i$
		\State $right\_tab[j-1]$ $\gets$ $right\_tab[j]$ $\times$ $voterPKs[i-1]$
		\EndFor
	\end{algorithmic}
	Compute the current batch of MPC keys:
	\begin{algorithmic}[1]
		\setcounter{ALG@line}{5}
		\For{$i \gets start$ to $end$}
		\State $act\_left \gets act\_left \times voterPKs[i-1]$
		\EndFor
		\State $mpc\_keys[i] \gets act\_left \div right\_tab[i] \;\mathrm{mod}\; mpc\_batch$
	\end{algorithmic}
\end{algorithm}

\section{Evaluation}
\label{sec:evaluation}
To evaluate the scalability of \name, we created an implementation of the protocol, which builds on BBB-Voting~\cite{bbbvoting}. 
We used the Truffle framework and Solidity programming language to implement the smart contract part and Javascript for the client API of all other components.
We also utilized the Witnet library~\cite{witnet-ecc-solidity} for on-chain elliptic curve operations on the standardized \emph{Secp256k1} curve~\cite{sec2002}.
Although Solidity was primarily intended for Ethereum and its Ethereum Virtual Machine (EVM), we have not selected Ethereum for our evaluation due to its high operational costs and low transactional throughput.
However, there are many other smart contract platforms supporting Solidity and EVM, out of which we selected Gnosis~\cite{xdai} and Harmony~\cite{harmony} due to their low costs and high throughput.

\begin{figure*}[t]
	\centering
	\begin{minipage}[t]{.49\textwidth}
		\includegraphics[width=\columnwidth]{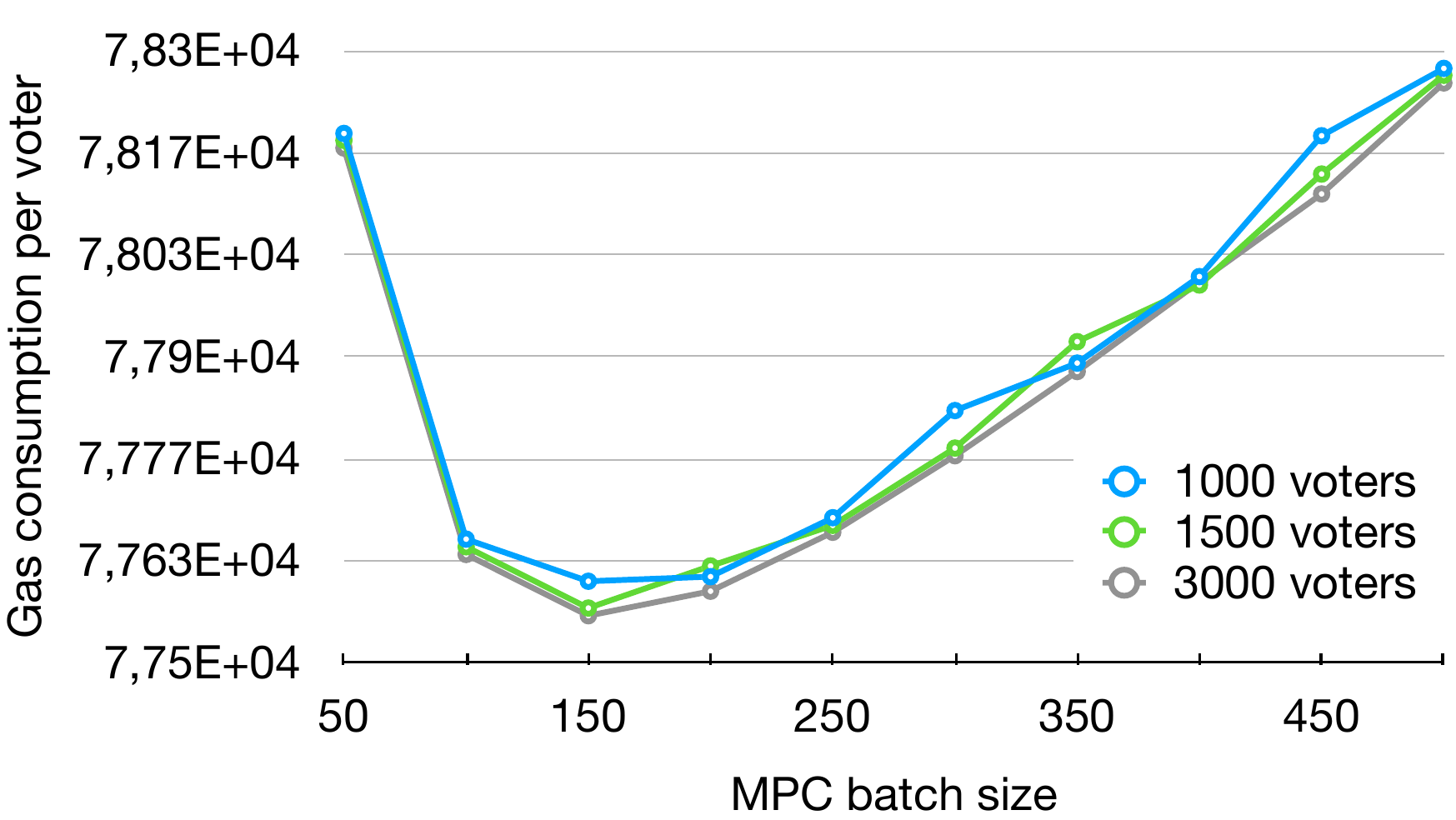}
		%		\vspace{-0.3cm}
		\caption{The cost [gas] per voter of the MPC key computation based on the chosen batch size.}
		\label{fig:mpc_batch}
	\end{minipage}\hfill
	\begin{minipage}[t]{.48\textwidth}
		\includegraphics[width=\columnwidth]{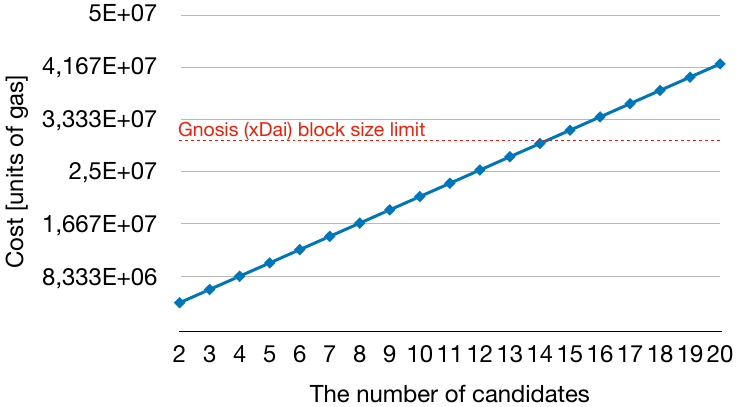}
		%		\vspace{-0.3cm}
		\caption{The cost [gas] of casting a vote based on the number of candidates.} 
		\label{fig:vote-cast}
	\end{minipage}
\end{figure*}

\myparagraph{\textbf{MPC batch size}}
The MPC keys in the Pre-Voting phase are computed in batches (see \autoref{ssec:optim}).
In detail, there is a pre-computed value available for the first voter in each batch.
This results in a trade-off between high overhead cost (in terms of units of gas) when selecting a small batch size resulting in many transactions and high execution cost due to utilizing fewer pre-computed values with larger batch sizes.
\autoref{fig:mpc_batch} illustrates how the batch size affects the cost of the computation per voter.

\myparagraph{\textbf{The number of candidates}}
The number of candidates our voting system can accommodate remains limited.
This is due to the block gas limit of a particular platform.
In detail, we can only run voting with a candidate set small enough so that the vote-casting transaction does not exceed the underlying platform's block gas limit.
Such transaction must be accompanied by a NIZK proof of set membership (i.e., proof that the voter's encrypted choice belongs to the set of candidates), and the size of the candidate set determines its execution complexity.
\autoref{fig:vote-cast} illustrates this dependency.
Our experiments show the proposed system can accommodate up to 38 candidates on Harmony and only 14 candidates on Gnosis.

\begin{figure*}[t]
	\centering
	\begin{minipage}[b]{.32\textwidth}
		\includegraphics[width=\columnwidth]{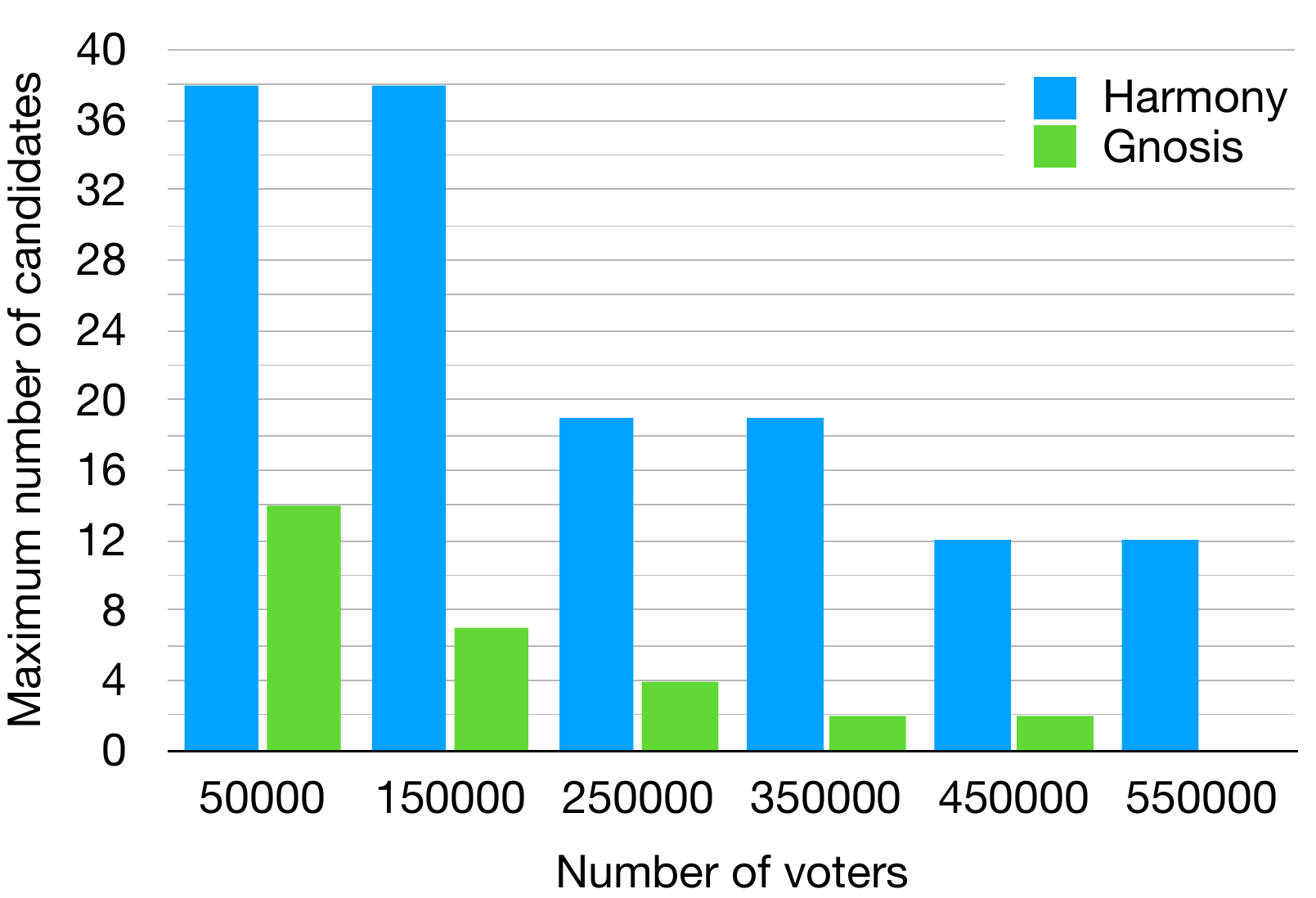}
		\caption{Maximum number of candidates (5-day voting phase).} 
		\label{fig:cand-vote}
	\end{minipage} \hfill
	\begin{minipage}[b]{.65\textwidth}
		\subfloat[On Gnosis.\label{fig:max_vot_xdai}]{	
			\includegraphics[width=0.47\columnwidth]{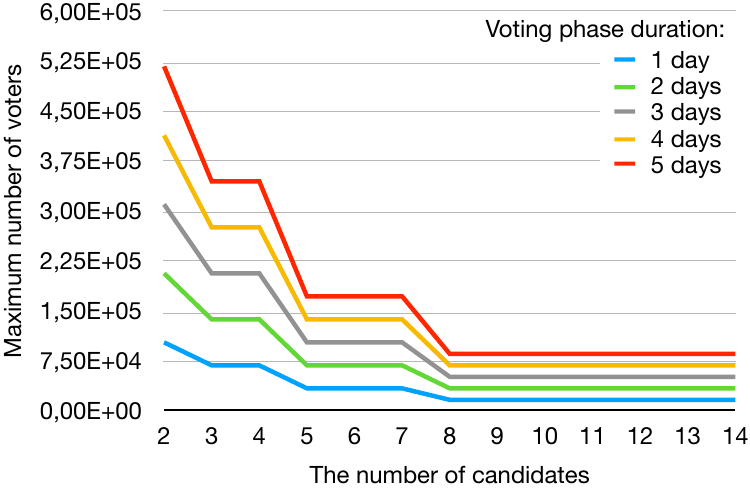}
		}
		\subfloat[On Harmony.\label{fig:max_vot_harmony}]{	
			\includegraphics[width=0.47\columnwidth]{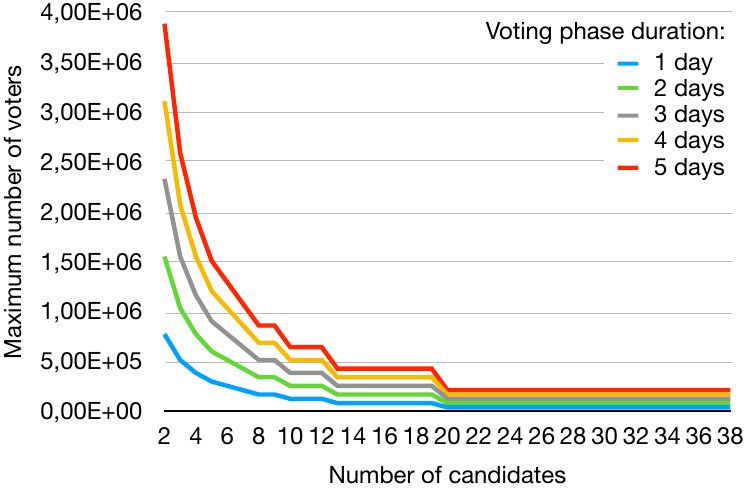}	
		}
		\caption{The maximum number of voters \name can accommodate.}
		\label{fig:max_voters}
	\end{minipage}
\end{figure*}

\myparagraph{\textbf{The total number of participants}}
The time period over which the voters can cast their ballots typically lasts only several days in real-life elections.
The platform's throughput over a restricted time period and the high cost of the vote-casting transactions result in a trade-off between the number of voters and the number of candidates.
We evaluated the limitations of the proposed voting protocol on both Harmony and Gnosis, as shown in \autoref{fig:cand-vote} and \autoref{fig:max_voters}.
Note that in these examples, we considered only the voting phase of the protocol to be time-restricted.
We determined that with two candidates, the proposed system can accommodate $\sim$1.5M voters over a 2-day voting period or up to 3.8M voters over a 5-day voting period on the Harmony blockchain.
However, with 38 candidates (maximum possible on Harmony), over a 5-day voting period, only 216K voters can participate.

\section{Security Analysis and Discussion}
\label{sec:discussion}
We discuss the properties and scenarios affecting the security of \name.

\myparagraph{\textbf{Privacy}}
Within each voting group, \name maintains perfect ballot secrecy.
The adversary, as defined in \autoref{ssec:sys-model}, cannot reveal a participant's vote through a collusion of all remaining participants since adversary can control at most $n-2$ participants.
The privacy of votes can be violated only if all participants in a voting group vote for the same candidate.
However, this is a natural property of voting protocols, which output the tally rather than only the winning candidate.
\name mitigates this problem by implementing transaction batching that eliminates bottlenecks affecting the size of the voting groups.
This allows the authority to maintain a sufficiently large size of the voting groups to lower the probability of a unanimous vote within the groups.
We refer the reader to the work of Ullrich \cite{ullrich2017risk} that addresses the issue of unanimous voting and the probability of its occurrence.

\myparagraph{\textbf{Deanonymization \& Linking Addresses}}
In conventional blockchains, the network-level adversary might be able to link the participant's address with her IP address. 
Such an adversary can also intercept the participant's blinded vote; however, she cannot extract the vote choice due to the privacy-preserving feature of our voting protocol.
Therefore, even if the adversary were to tie the IP address to the participant's identity, the only information it would be able to obtain is whether she has voted.
Nevertheless, to prevent the linking of addresses, the participant can use VPNs or anonymization services such as Tor.

\myparagraph{\textbf{Re-Voting}}
It is important to ensure that no re-voting is possible, which is to avoid any inference about the final vote of a participant in the case she would reuse her ephemeral blinding key to change her vote during the voting stage.
Such a re-voting logic can be easily enforced by the smart contract, while the user interface of the participant should also not allow re-voting.
Also, note that ephemeral keys are one-time keys and thus are intended to use only within one instance of e-voting protocol to ensure the security and privacy of the protocol.
If a participant were to vote in a different instance of e-voting, she would generate new ephemeral keys.

\myparagraph{\textbf{Forks in Blockchain}}
Since our protocol does not contain any two-phase commitment scheme with revealed secrets, its security is not impacted by accidental or malicious forks. 
Temporary forks do not impact the voting stage as well since the same votes can be resubmitted again by client interfaces.

\myparagraph{\textbf{Self-Tallying Property}}
We note that the self-tallying property only holds within each voting group, not for the voting instance as a whole.
The main contract aggregates the booth tallies, not the actual blinded votes.
However, the integrity of this step is enforced by the smart contract and is fully verifiable since all booth tallies are publicly available on the blockchain.

\myparagraph{\textbf{Verifiability}}
\name achieves both individual and universal verifiability.
By querying the booth contract, each voter can verify her vote has been recorded. 
Each voter, as well as any interested third party, can verify the booth tally since it satisfies the self-tallying property, i.e., the \autoref{eqn:eq3} would not hold should any vote be left out.
Any party can verify the final tally aggregated by the main contact by querying the booth contracts to obtain individual booth tallies.

\myparagraph{\textbf{Platform-Dependent Limitations}}
Although our system itself does not limit the number of participants, the required transactions are computationally intensive, which results in high gas consumption.
Therefore, large-scale voting using our system might be too demanding for the underlying smart contract platform.
As a potential solution, public permissioned blockchains dedicated to e-voting might be utilized. 

\myparagraph{\textbf{Adversary Controlling Multiple Participants in Fault Recovery}}
One issue that needs to be addressed in the fault recovery is an adversary controlling multiple participants and letting them stall one by one in each fault recovery round.
Even though the fault recovery mechanism will eventually finish with no new stalling participants, such behavior might increase the costs paid by remaining participants who are required to submit counter-party shares in each round of the protocol.
For this reason, similar to the voting stage, we require the fault recovery stage to penalize stalling participants by losing the deposit they put into the smart contract at the beginning of our protocol.
On the other hand, the adversary can cause a delay in the voting protocol within a particular booth, which, however, does not impact other booths. 
To further disincentivize the adversary from such behavior, the fault-recovery might require additional deposits that could be increased in each round by some constant, while all deposits could be redeemed at the tally stage.

\myparagraph{\textbf{Tally computation}}
Tallying the results in individual booths requires an exhaustive search for a solution of \autoref{eqn:eq3} with $\binom{n+k-1}{k-1}$ possible solutions \cite{HaoRZ10}, where $n$ is the number of votes and $t$ is the number of candidates.
Therefore, the authority should select the size of the voting groups accordingly to the budget and available computational resources (see \cite{bbbvoting} for experimental evaluation).

\section{Related Work}
\label{sec:relatedwork}
In this section, we provide a brief survey of existing e-voting solutions.

Several protocols have been proposed, focusing on ensuring the vote's privacy rather than breaking the map between the voter and her ballot.
Cohen and Fisher~\cite{Cohen1985} proposed a verifiable voting scheme where the participants cannot unveil the votes.
However, the election authority in this scheme has the ability to read any vote.
Cohen~\cite{Cohen86} provided an extension to this scheme, where the function of authority is distributed among an arbitrary number of \emph{tellers}. At least one honest \emph{teller} is sufficient to ensure the privacy of the votes. 

Several other works based on the approach from \cite{Cohen1985}, such as \cite{cgs97,Benaloh1986,cramer1996multi,Schoenmakers99}.
The multi-authority protocol proposed by Cramer et al.~\cite{cgs97} employs the ElGamal cryptosystem to guarantee vote privacy.
This protocol can tolerate malicious behavior of a constant fraction of authorities.
Baudron et al.~\cite{Baudron2001} focused on multi-candidate elections with hierarchical levels of authorities.

Kiayias and Yung~\cite{Kiayias2002} introduced a new voting paradigm with properties they defined -- self-tallying, perfect ballot secrecy, and dispute-freeness.
The protocol presented by Groth~\cite{Groth2004} improved the computational complexity of \cite{Kiayias2002} but required more rounds of computation as a trade-off.
Hao et al.~\cite{HaoRZ10} further improved this approach and created a 2-round self-tallying voting scheme.
Khader et al.~\cite{KhaderSRH12} proposed a variant of \cite{HaoRZ10} that also ensures fairness and robustness.

Protocols based on \cite{Cohen1985} and \cite{Kiayias2002} as well as other approaches (\cite{lueks2020voteagain,RyanBHSX09,Adida08,cortier2019belenios,clarkson2008civitas}) require a public bulletin board (PBB), defined as a broadcast channel with memory.
As defined, PBB is not affected by denial-of-service attacks and allows each participant to write solely in her designated section in an append-only manner. 
To achieve these properties in practice, Cramer et al.~\cite{cgs97} suggest implementing PBB as a set of replicated servers running a Byzantine agreement protocol.
The introduction of blockchain technology brought a suitable solution for a practical instantiation of PBB since it offers the required properties of immutability and accessibility.

McCorry et al.~\cite{McCorrySH17} were the first to implement the self-tallying scheme using smart contracts on Ethereum. However, their system, the Open Vote Network, is only suitable for a small-scale (boardroom) voting.
Venugopalan et al.~\cite{bbbvoting} presented BBB-Voting, also a boardroom voting protocol, but with several improvements in contrast to  OVN.
BBB-voting~\cite{bbbvoting} supports multiple candidate choices and provides cost-optimized implementation on Ethereum.
Seifelnasr et al.~\cite{SeifelnasrGY20} improved the scalability of \cite{McCorrySH17} by reducing the storage requirements and delegating the tally computation to an off-chain entity.

Besides the implementations of the self-tallying approach, other blockchain-based voting systems have been proposed, such as Zhang et al.~\cite{Zhang18}, Alvi et al.~\cite{alvi2020privacy}, Dagher et al.~\cite{icissp:DMMM18} (BroncoVote), Killer et al.~\cite{killerprovotum} (Provotum) or Zhang et al.~\cite{zhang2020chaintegrity} (Chaintegrity).
Blockchain-based voting was also criticized by Park et al.~\cite{park2020going} for bringing additional security issues rather than improvements.

\section{Conclusion}
\label{sec:conclusion}
In this paper, we present a scalable self-tallying blockchain-based voting protocol.
We made an implementation of the protocol and evaluated its performance on two EVM-compatible platforms -- Gnosis and Harmony.
We showed that our protocol is fully scalable to accommodate large-scale voting, with the only limitation being the throughput of the underlying blockchain platform. 
Our experiments show that our system can run voting with millions of participants on a sufficiently fast blockchain (e.g., Harmony).

In our future work, we will focus on replacing the NIZK proofs of the voting phase with zk-SNARKs to achieve a constant complexity of vote casting independent of the number of candidates.
Further, we intend to investigate the techniques used to increase the throughput of smart contract platforms (e.g., sharding) and analyze the impact of these approaches on the security properties and scalability of \name.

\bibliographystyle{splncs04}
\bibliography{ref}
%
%\fi

\end{document}